# User Authentication on Earable Devices via Bone-Conducted Occlusion Sounds

Yadong Xie, *Graduate Student Member, IEEE*, Fan Li, *Member, IEEE*, Yue Wu, and Yu Wang, *Fellow, IEEE*

*Abstract*—With the rapid development of mobile devices and the fast increase of sensitive data, secure and convenient mobile authentication technologies are desired. Except for traditional passwords, many mobile devices have biometric-based authentication methods (e.g., fingerprint, voiceprint, and face recognition), but they are vulnerable to spoofing attacks. To solve this problem, we study new biometric features which are based on the dental occlusion and find that the bone-conducted sound of dental occlusion collected in binaural canals contains unique features of individual bones and teeth. Motivated by this, we propose a novel authentication system, TeethPass⁺, which uses earbuds to collect occlusal sounds in binaural canals to achieve authentication. First, we design an event detection method based on spectrum variance to detect bone-conducted sounds. Then, we analyze the time-frequency domain of the sounds to filter out motion noises and extract unique features of users from four aspects: teeth structure, bone structure, occlusal location, and occlusal sound. Finally, we train a Triplet network to construct the user template, which is used to complete authentica- tion. Through extensive experiments including 53 volunteers, the performance of TeethPass⁺ in different environments is verified. TeethPass⁺ achieves an accuracy of 98.6% and resists 99.7% of spoofing attacks.

## I. Introduction

NOWADAYS, mobile devices are becoming powerful with a large storage capacity. They are often used to process sensitive information (e.g., private documents edit, health information record, and online payment). However, the leakage of user privacy data is increasingly serious. According to a survey from Cisco [1], 89% of users care about privacy data security, and 79% of them are willing to act to protect it. Another report from IBM [2] shows that personally identifiable information, such as login data, fingerprints, and voice, is the most frequently lost or stolen type of data. Thus, it is necessary to study a reliable and convenient authentication system.

To prevent the leakage of user privacy data, many authentication methods are adopted on mobile devices (e.g., PIN code, unlock pattern, and fingerprint). But the token used by these methods is susceptible to being inferred or stolen. Specifically, the PIN code and unlock pattern are the most popular authentication methods, but they are vulnerable to attacks [3] and require tedious input by users. Besides, many types of biometric features are studied for user authentication, such as fingerprint [4], [5], voiceprint [6], [7], and face recognition [9], [10], which are also adopted on commercial systems (e.g., Apple Touch ID [11], TD VoicePrint [12], and Amazon Rekognition [13]). However, these methods are vulnerable to replay attacks. For instance, an attacker can record the victim's face or voice, then replay the records to spoof the authentication system. Even fingerprints can be stolen through photos and made into fingerprint film for attacks. Recently, more types of behavioural [14], [15], [16] and biometric [17], [18], [19] features are leveraged to enhance the security of mobile authentication. PressPIN [15] is an enhanced PIN on mobile devices by sensing the structure-borne propagation of sounds to estimate the pressure on the screen. FINAUTH [16] extracts fingertip-touch characteristics by mobile devices to complement fingerprint authentication. SmileAuth [17] adopts the image of users' dental edge for authentication. However, it requires the user to hold a phone towards the mouth, which is inconvenient and works in a limited scenario. EarEcho [19] uses features of human ear canal and assesses acoustic features of in-ear sound for authentication. But it is susceptible to interference from environments, such as position shift of device.

Motivated by the above limitations, we design a secure, convenient, and reliable user authentication method, TeethPass⁺, based on the bone-conducted sound [20], [21], [22] of dental occlusion. Specifically, when a user occlude teeth, the occlusal sounds are absorbed, reflected, and dispersed by the skull and then transmitted to ear canals. Thus, the received sounds present individual differences due to the unique density and elasticity properties of skull. With this characteristic, the occlusal sound (received in ears) can be used for authentication. Additionally, wireless earbuds are used widely in recent years. A survey [23] shows that the number of wireless earbuds in 2024 will reach 520 million, and many users report that they tend to wear earbuds all day. Most of all, commercial earbuds (e.g., Apple AirPods Pro, Sony WF-1000XM4, and Bose QuietComfort)

Manuscript received 12 September 2022; revised 19 October 2023; accepted 12 November 2023. Date of publication 28 November 2023; date of current version 11 July 2024. The work of Fan Li is partially supported by the National Natural Science Foundation of China (NSFC) under Grants 62372045 and 62072040. *(Corresponding author: Fan Li.)*

This work involved human subjects or animals in its research. Approval of all ethical and experimental procedures and protocols was granted by the Ethics Committee of Beijing Institute of Technology under Application No. 2023081, and performed in line with the principles of the Declaration of Helsinki.

Yadong Xie and Fan Li are with the School of Computer Science and Technology, Beijing Institute of Technology, Beijing 100089, China (e-mail: ydxie@bit.edu.cn; fli@bit.edu.cn).

Yue Wu is with the School of Software, Tsinghua University, Beijing 100089, China (e-mail: ywu92@mail.tsinghua.edu.cn).

Yu Wang is with the Department of Computer and Information Sciences, Temple University, Philadelphia, PA 19122 USA (e-mail: wangyu@temple.edu).

Digital Object Identifier 10.1109/TDSC.2023.3335368

have inward-facing microphones to collect the sound in ear canals for noise reduction. These principles inspire the basic idea of TeethPass+: to use the inward-facing microphones of earbuds to capture bone-conducted sounds of dental occlusion, and then to extract the unique biometric features from these sound for user authentication.

Despite its simple idea, three major challenges underlie the design of TeethPass+. First, although the inward-facing microphone is very close to the ear canal, it still can record air-conducted noises (e.g., speaking and environment noise). Besides, some daily activities also generate bone-conducted noises (i.e., chewing, gulping, speaking, and walking). So, how to distinguish the bone-conducted occlusal sounds from the collected noisy acoustic signal is the first challenge. Second, TeethPass+ relies on unique bone-conducted occlusal sounds for user authentication. Thus, how to extract unique biometric features contained in the bone-conducted sounds to characterize the skull biometrics and occlusal location diversity is the second challenge. Finally, it is necessary for an authentication system to collect as little registration data as possible to achieve user-friendliness. Thus, we need to achieve an efficient authentication model with limited registration data.

To address the above challenges, we first analyze occlusion events and typical daily activities. We find that occlusion events have much shorter duration than chewing and gulping. Moreover, the frequency of occlusal sounds ranges from 100 Hz to 2.5 kHz, while the bone-conducted sounds of walking and speaking have frequencies mainly below 300 Hz. Therefore, occlusion events can be distinguished from daily activities in terms of duration, frequency range, and energy distribution. Then, to characterize a user's skull biometrics and occlusion location, we extract 4 biometric features, including the teeth structure feature related to occlusal contact surface (i.e., teeth enamel and cusp), the dispersion feature [24] related to the structure of bone, the propagation delay related to the occlusal location, and the acoustic feature related to the bone-conducted sound. Finally, a Triplet network [25] based authentication scheme is designed to identify users accurately. We apply data augmentation methods (i.e., time warping and time-frequency masking) to the limited training data. Combined with incremental learning [26], we can quickly update the parameters of the network to authenticate the newly registered user.

We implement TeethPass+ by using 3 kinds of earphones with inward-facing microphones. We recruit 53 participants (34 males and 19 females) and ask them to put on earphones for occlusion in diverse scenarios. We also simulate 3 types of attacks to test the anti-attack ability of our system. The results demonstrate that TeethPass+ is accurate in different environments, and can resist various spoofing attacks.

Our contributions are summarized as follows:
- We propose a novel authentication system, TeethPass+, which uses earbuds to collect bone-conducted sounds of dental occlusion in binaural canals. To the best of our knowledge, we are the first to sense occlusal sounds by earbuds for authentication.
- We propose effective methods to filter out interferences of daily actions. We also design 4 unique biometric features for authentication, including structure features of teeth and bone, location features of occlusion, and acoustic features of occlusal sounds.
- We build an authentication scheme based on a Triplet network. And we combine the scheme with incremental learning, which can quickly update the parameters of the network for authenticating newly registered users.
- We evaluate TeethPass+ by using 3 prototypes in different application scenarios. The results show that TeethPass+ can authenticate users with an average accuracy of 98.6%, and resist 99.7% of attacks.

## II. RELATED WORK

In this section, we review 3 kinds of biometrics-based authentication systems related to TeethPass+.

*Voice-Based Authentication:* Among various biometric-based authentication methods, voiceprint is one of the most commonly used biometrics for authentication. But traditional voiceprint-based methods [6], [7], [8] are vulnerable to replay attacks. To improve security, VoicePop [27] leverages pop noises that are produced when a user is breathing and can be hardly maintained in records to achieve authentication. LipPass [18] extracts unique features from users' speaking lips leveraging active acoustic sensing on smartphones. EarPrint [28] aims to extend voiceprint by building on body sounds that transmit from the throat to the ear for authentication. SoundID [29] is based on dynamic acoustic fingerprints and ambient sound, which are derived from both the login device and the enrolled phone for each authentication. However, these systems require users to speak and are not suitable in some environments (e.g., library and conference room).

*Teeth-Based Authentication:* Teeth biometrics, such as size, shape, and edge envelope, are intrinsically unique among individuals [30]. SmileAuth [17] extracts dental edge features by the smartphone to capture a few images from different camera angles for authentication. An authentication approach [31] utilizes the contour information of teeth to extract coarse-grained features and further employs voice data to improve the accuracy. But these methods are sensitive to light and vulnerable to replay attacks. In addition to images, occlusal sounds can also be used as biometrics. ToothSonic [32] leverages the toothprint-induced sonic effect produced by a user performing teeth gestures for user authentication. But it lacks the processing of motion noise, which limits the use scene. BiLock [33] extracts features from the sounds generated by a user's occlusion, which are recorded by the built-in microphone of a smartphone placed close to the user's lips. However, it only uses air-conducted occlusal sounds for authentication, so its principle is more similar to traditional voiceprint-based authentication, which makes it vulnerable to spoofing attacks.

*In-Ear Authentication:* Recently, the development of smart earbuds provides a new way for user authentication. EarEcho [19] extracts the features by emitting sounds from the earphone. The sounds are reflected through the ear canal which can be recorded by inward-facing microphones. An authentication system [34] utilizes the microphone-integrated earphone to capture the static ear canal geometry. It extracts features of the reflected signals from a ear canal to distinguish different users. EarDynamic [35] makes earbuds emit an inaudible signal to probe the ear canal. Then the signal reflected from the ear canal are captured by the inward-facing microphone that can be further utilized to extract the deformation of the ear canal. However, most of these methods require earphones to emit ultrasound, and may impair the health of users if they are exposed to the ultrasound with high volume [36].

Different from these works, TeethPass+ uses earbuds to collect bone-conducted sounds of dental occlusion in binaural canals to achieve authentication, which can effectively resist both replay attacks and mimic attacks. Furthermore, the sound of dental occlusion is more imperceptible and unobtrusive than voice, so it is more socially acceptable. As an extension to our conference paper [37], TeethPass+ comprehensively analyzes the motion noise that may affect the system performance. TeethPass+ further explores deeper biometric features related to occlusal contact surface. We also adopt a new feature extraction method to extract more fine-grained acoustic features and designs a more innovative Triplet network to authenticate users, which achieves higher authentication accuracy and better robustness against many interference factors.

## III. PRELIMINARY

In this section, we first introduce the theoretical analysis of occlusal sound, then exploit the feasibility of using bone-conducted sounds of occlusion for authentication, finally analyze potential attack models.

### A. Theoretical Analysis of Occlusal Sound

Bone-conducted sounds of dental occlusion have unique biometric features. On the one hand, previous research [38] shows that dental structure is stable over time, and even a single tooth of a person is unique. The single tooth features include enamel thickness, enamel rod patterns, cusp morphology, and so on [39]. Data from a study [40] suggest that differences in enamel thickness and dentine area exist between different sexes and populations, such as significantly greater average enamel thickness in females. Enamel rod patterns refer to different patterns of enamel rod endings on tooth surfaces. These are also called as tooth prints and they help in personal identification in forensic odontology [41]. And the number of cusps of the molar can range from 4 to 7 for different individuals [42]. Occlusion [43] refers to the action between the upper and lower teeth when they approach each other, so the occlusal sound has unique features for an individual.

On the other hand, the occlusal sound passes through the maxilla, mandible, and zygoma, finally arrives at the auditory meatus. These bones have unique physical features [44], such as shape, bone-muscle ratio, and density, which lead to the unique dispersion, absorption, and reflection of occlusal sounds. Given the theoretical analysis of bone-conducted sounds of occlusion, we conduct feasibility experiments to verify the uniqueness of occlusal sounds.

### B. Feasibility Study

To verify the feasibility of using bone-conducted sounds of occlusion to complete authentication, we collect occlusal sounds of 7 volunteers (4 males and 3 females, aged 18 to 46). Before collection, we explain the principle of TeethPass+ to them. Then, they are asked to find 1-3 comfortable occlusal locations and practice occlusion a few times. We use a pair of inward-facing microphones in ears to record bone-conducted sounds. All sounds are collected with a 48 kHz sampling rate. As the feasibility study, the experiments are conducted under the low noise environments.

*Inter-User Study:* First, we ask different volunteers to use teeth in similar location for occlusion. The Continuous Wavelet Transform (CWT) results of their bone-conducted sounds are shown in Fig. 1(a)–(c). The results show noticeable individual differences refer to duration time, frequency range, and energy distribution. Besides, the occlusal sounds collected from all volunteers are visualized in Fig. 1(d) with the t-distributed Stochastic Neighbor Embedding (t-SNE) [45] method, which is suitable to reduce the dimension of high-dimensional datasets. We can find that even if the locations of occlusion are similar, the occlusal sound of each volunteer shows a unique and consistent pattern, which presents that the individual difference in skull biometric features can be captured by the bone-conducted sounds.

*Intra-User Study:* Then, we study the bone-conducted sounds of a volunteer when occluding with teeth at different locations. Fig. 2(a)–(c) show the results of CWT on the bone-conducted sounds of three locations (i.e., left, middle, and right teeth). We can see that when the same volunteer occludes different teeth, the time-frequency domain features are different. Fig. 2(d) shows the bone-conducted sounds after visualization of 2 volunteer at 3 locations. Since different teeth have different structure features, and the propagation paths of sound to the ear canal are also different, the occlusal sounds generated by teeth at different locations are quite diverse even for the same volunteer. Thus, the distance between intra-user features is large. In *SYSTEM DESIGN* (Section IV), we adopt more effective biometric features and authentication schemes. Thus, compared to the feasibility study, our designed approach can better distinguish different users.

### C. Attack Model

Here we list two main types of spoofing attacks for off-the-shelf authentication, i.e., mimic attack and replay attack. Furthermore, we also consider an extreme scenario in which the spoofers implement mimic attacks and replay attacks at the same time to achieve hybrid attacks.

*Mimic Attack:* In order to attack a voice-based authentication, spoofers first observe the way of speaking when legitimate users login, then practice to mimic the tone, speed, and pronunciation to perform the mimic attack. If spoofers attempt to conduct mimic attacks on our system, they first need to know which teeth and how much force legitimate users use to occlude when using our system. Then, they can wear users' earbuds and mimic the teeth occlusion of the user to spoof our system.

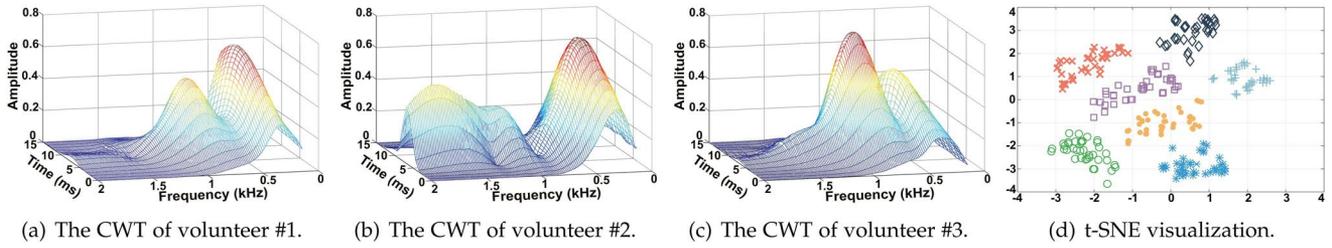

Fig. 1. Inter-user study.

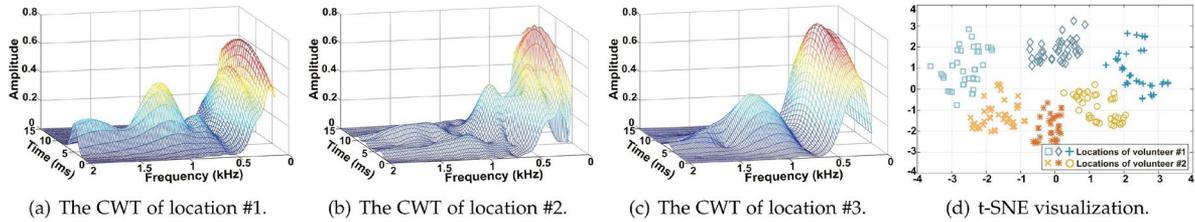

Fig. 2. Intra-user study.

*Replay Attack:* For teeth-image-based authentication, spoofers can use high-definition cameras to secretly take photos or videos of legitimate users' teeth and replay them to spoof the authentication system. The voice-based authentication requires users to make sounds, so there is a high probability that spoofers eavesdrop and record the voice of legitimate users. Then the spoofers can spoof the authentication system by playing back the recorded voice. For our system, spoofers may collect air-conducted sounds of dental occlusion at a location very close to the legitimate users, and replay them to perform replay attacks.

*Hybrid Attack:* We also consider an extreme situation, that is, spoofers can not only collect the air-conducted sounds of dental occlusion but also know the occlusal location and force of legitimate users during authentication. So they can mimic the occlusion action of legitimate users while playing the recorded occlusal sounds by bone conduction speakers in spoofers' mouths.

## IV. SYSTEM DESIGN

In this section, we first present the system overview of TeethPass+ and then detail the behind techniques.

### A. System Overview

Fig. 3 shows the architecture of TeethPass+, which can be divided into registration phase and login phase.

In the registration phase, TeethPass+ aims to collect registration data from a user and generate the user template that contains biometric features. Before collection, the user needs to find a comfortable occlusal location and practice occlusion a few times. During collection, the user occludes teeth several times, and occlusal sounds are recorded by a pair of inward-facing microphones in the ears. The recorded sounds first go through *Air Noise Removal*, including normalization, noise reduction, and frequency selection, to reduce air-conducted ambient noises. Then, TeethPass+ performs *Event Detection* to segment each bone-conducted sound event. Specifically, we calculate spectrum variance to capture the energy in different frequency bands of sound, then adopt a double-threshold method to detect the start and end points of each event. Since the sound may reach ears at different times, we unify the length of each event received by the two microphones. Some daily actions (i.e., chewing, gulping, speaking, and walking) also produce bone-conducted sounds and can be captured by *Event Detection*, so the events that contain these daily actions are removed in *Motion Removal*. We analyze each event's duration and power spectrum density (PSD) to distinguish occlusion from the daily actions. For each occlusal event, four biometric features are extracted, including the teeth structure feature related to occlusal contact surface, the bone structure feature related to physical properties of bone, the location feature presented as occlusal sound delay, and the acoustic feature related to bone-conducted sound. Finally, the teeth structure and acoustic features are used to train a Triplet network. The trained network and two other kinds of features are stored in the user template.

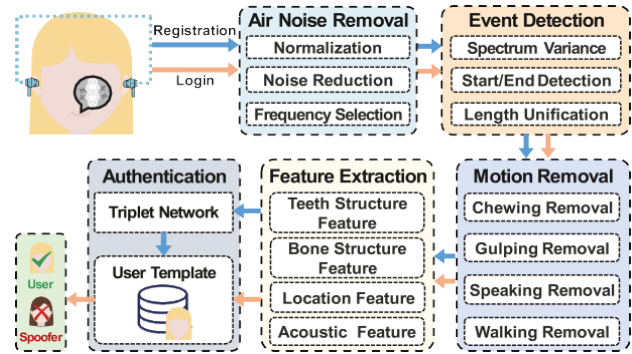

Fig. 3. System architecture of TeethPass+.

In the login phase, the bone-conducted sounds of a user is recorded by a pair of inward-facing microphones in the ears. Then, through *Air Noise Removal*, *Event Detection*, *Motion Removal*, and *Feature Extraction*, the four biometric features of the login user are obtained. Finally, we compare the similarities between them and the template of each registered user to achieve user authentication.

### B. Air Noise Removal

When a user puts on earbuds, TeethPass+ starts to monitor sounds in ear canals in real-time. Although the inward-facing microphone faces the ear canal, it may still record air-conducted ambient noises, such as human voice and road noise. Thus, the raw recorded sounds need to be processed to filter out the interference of air-conducted ambient noises. To ensure real-time, we add a sliding window to the sounds. After observation, we find that the duration of the occlusal sound is usually between 10 ms and 20 ms, so the length of the sliding window is 50 ms and it slides 10 ms each time. The sound in each window is a frame.

*Normalization:* There may be a slight difference in the tightness and angle of the earbuds each time the user puts on them, so the volume of sounds recorded by inward-facing microphones may be unstable. Before noise reduction, we use normalization to process each frame. The most common method is peak normalization [46], [50], [51], which adjusts the sounds based on the highest volume level in each frame. However, it leads to the problem that the average volume is inconsistent across frames. So we use another normalization based on loudness, which adjusts the average volume of each frame to a desired volume. We set the desired volume to −24 dB, which is the same as the standard loudness recommended by Advanced Television Systems Committee ATSC [52].

*Noise Reduction:* Then, we reduce air-conducted ambient noises in each frame to improve the signal-to-noise ratio. Considering the limited computing capability of mobile devices, we adopt power spectral subtraction which has the advantages of small computation and high processing speed. Suppose the audio signal of a frame is $x(m)$, and $X(k)$ denotes the fast Fourier transform (FFT) results of $x(m)$. The amplitude $|\hat{X}(k)|$ after spectral subtraction can be calculated by

$$\hat{X}(k)^2 = \begin{cases} |X(k)|^2 - aD(k), & |X(k)|^2 \geq aD(k), \\ \beta D(k), & |X(k)|^2 < aD(k), \end{cases} \quad (1)$$

where $a$ and $\beta$ are constants, representing the over-subtraction factor and the spectral floor parameter, respectively. $D(k)$ denotes the amplitude of environmental noise recorded by outward-facing microphones. Through inverse FFT of $|\hat{X}(k)|$, we get the audio signal $\hat{x}(m)$ after noise reduction. It is obvious that power spectral subtraction can reduce the impact of different noise magnitudes caused by the different environments.

Since we adopt outward-facing and inward-facing microphones to complete spectral subtraction, we further study whether air-conducted occlusal sounds affect the result, and thus interfere with bone-conducted occlusal sounds. Fig. 4(a) shows the air-conducted occlusal sound received by the outward-facing microphone. We can see that the energy of the sound conducted by air is much lower than that conducted by bone (Fig. 4(b)), since the air-conducted sounds attenuate faster than bone-conducted sounds [53]. It proves that the spectral subtraction has no effect on bone-conducted occlusal sounds.

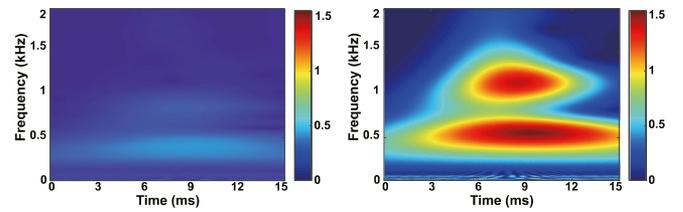

(a) Air-conducted sound. (b) Bone-conducted sound.

Fig. 4. Occlusal sound spectrums of air and bone conduction.

*Frequency Selection:* Finally, we use a band-pass filter to select the frequency band of occlusal sounds. Through our observations, we find that most bone-conducted sounds of occlusion range from 100 Hz to 2.5 kHz. Therefore, we adopt a Butterworth band-pass filter ranging from 100 Hz to 2.5 kHz for each frame. After filtering, we can eliminate other out-band interferences and prepare for *Event Detection*.

### C. Event Detection

After *Air Noise Removal*, the sound in each frame almost only contains bone-conducted sound. Then, we detect whether there is a occlusal like event in each frame and segment each possible occlusion event. A common detection method is based on Short Time Energy [26], [54], [55], which is widely used in speech recognition and motion detection. But we find that the energy of bone-conducted sound of occlusion varies greatly with frequency, while most other bone-conducted sounds are more evenly distributed in the spectrum. Thus, we divide the spectrum into several bands and study an detection method based on spectrum variance.

*Spectrum Variance Calculation:* First, we apply a sliding window with a length of 2.5 ms that slides 1 ms each time on each frame. The signal in the $i$-th window is $x_i(m)$, we divide the amplitude $|X_i(m)|$ obtained by the FFT into $q$ sub-bands evenly. And each sub-band is formed as

$$S_i(n) = \sum_{k=1+(n-1)p}^{1+(n-1)p+(p-1)} |X_i(k)|, n \in [1, q], \quad (2)$$

where $p$ is the number of frequency points in each sub-band. Then, the spectrum variance $D_i$ can be calculated by

$$D_i = \frac{1}{q-1} \sum_{k=1}^{q} \left( S_i(k) - \frac{1}{q} \sum_{s=1}^{q} S_i(s) \right)^2. \quad (3)$$

By analyzing (3), we find that the greater the fluctuation between the frequency bands, the greater the $D_i$. Fig. 5 shows the bone-conducted sounds of two occlusion and the corresponding

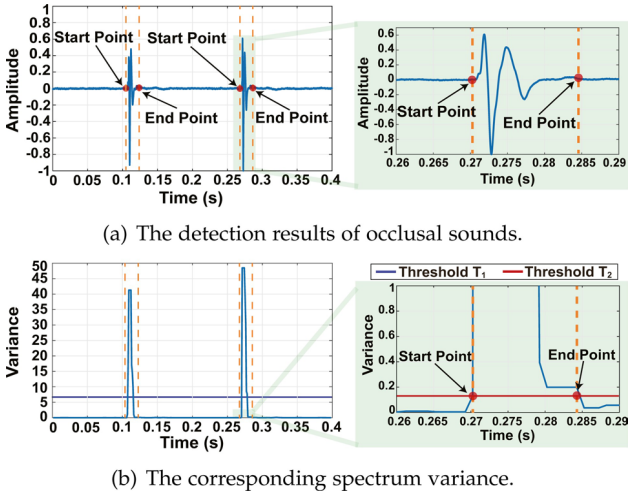

(a) The detection results of occlusal sounds.

(b) The corresponding spectrum variance.

Fig. 5. Process of event detection.

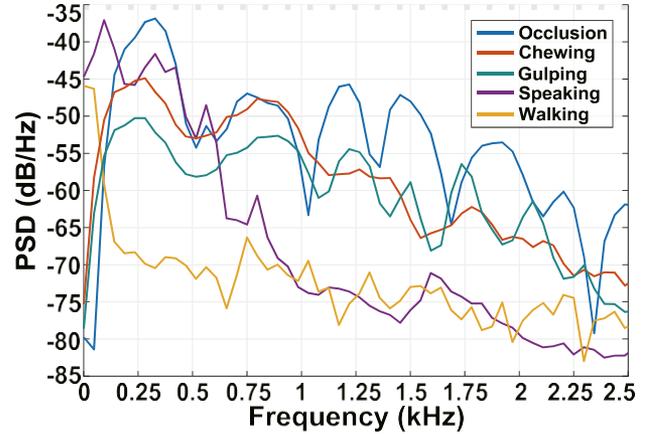

Fig. 6. Power spectrum density of five actions.

spectrum variance. It is shown that spectrum variance can be used to capture the occlusal sounds well.

*Start/End Detection:* Then, we adopt a double-threshold method [56] to detect the start point and end point of each event. Specifically, we first set a threshold $T_1$ on spectrum variance, and the segment which is larger than $T_1$ can be considered to contain an event. Then, another threshold $T_2$ ($T_2 < T_1$) is set to find the start and end points of the event. We search from the beginning of the segment to the left and find the first point that intersects with $T_2$ as the start point of the event. In the same way, we search from the ending of the segment to the right to find the end point of the event. We select the thresholds by analyzing the collected data, we use different threshold values to construct the system and evaluate the system performance. The thresholds for achieving the best system performance are selected as $T_1$ and $T_2$. Fig. 5 also shows the detection results of start points and end points on occlusal sounds and spectrum variance. We can see that the double-threshold method identifies start and end points precisely.

*Length Unification:* We collect bone-conducted sounds using a pair of inward-facing microphones and process the sounds of the two microphones separately. Since the user can choose the occlusal location arbitrarily, the occlusal sound conducts to the two ear canals in different paths, which causes the start and end points of the occlusal sounds received by the two microphones to be different. In order to facilitate *Feature Extraction* later, we unify the length of two events received by the two microphones. Specifically, we choose the smaller one of the two start points as the new start point of the two events, and the larger one of the two end points as the new end point of the two events. After length unification, each occlusal sound produces two events with the same length.

### D. Motion Removal

Although we filter out most of the air-conducted ambient noises in the *Air Noise Removal* and adopt an *Event Detection* method that meets the characteristics of bone-conducted sound of teeth occlusion, there are also some actions that produce bone-conducted sound similar to occlusal sound (i.e., chewing, gulping, speaking, and walking). These bone-conducted sounds are also extracted by *Event Detection*, so we need to remove these non-occlusion events from the detected events.

*Chewing Removal:* When users eat, they usually need to use their teeth to chew food. The chewing action is produced by the movements of the mandible. In this process, cutting, tearing, and grinding processes between the teeth and food can create collision and friction which induce bone-conducted sounds that are then passed into the auditory system. Experiment on foods with different hardness (e.g., rice, chips, apples, and nuts) demonstrates that different foods lead to distinct bone-conducted sounds. We calculate the power spectrum density (PSD) of different bone-conducted sounds, and find that the frequency range of chewing is similar to that of occlusion, as shown in Fig. 6. However, we note that the duration of a bone-conducted chewing sound is generally greater than 250 ms, while the duration of an occlusal sound is usually between 10 ms and 20 ms. So we can determine whether the event is chewing by analyzing the duration time.

*Gulping Removal:* The gulping consists of three phases, i.e., oral, pharyngeal, and esophageal phases. During the second and third phases, the movements of the larynx, hyoid bone, and different muscles in the pharynx can produce a gurgle sound which is due to the transmission of the food into the esophagus. The interactions between the food and the pharynx structure lead to bone-conducted gulping sounds that can be heard and recorded in the ear canal. By analyzing the PSD in Fig. 6, we can find that gulping sounds are very similar to bone-conducted occlusal sounds in frequency range and energy distribution. But in the time domain, the gulping sound has a longer duration, usually greater than 700 ms. Therefore, we can use the same method for chewing removal to remove gulping events.

*Speaking Removal:* The human voice is produced by the vibration of vocal cords, which can also be transmitted to the ear canal through bone conduction. Relevant research [57] and our experiment show that the base frequency of human voice is between 80 Hz and 300 Hz, while the frequency of occlusion is between 100 Hz and 2.5 kHz, as shown in Fig. 6. To detect whether the event is speaking, we calculate the energy ratio of the PSD of the 100 Hz-300 Hz to the PSD of the 100 Hz-2.5 kHz. If the energy ratio is greater than a threshold, we consider that the event is speaking. It should be noted that users can not authenticate while chewing, gulping, and speaking, so we just discards these events when they are detected.

*Walking Removal:* Authentication while walking is a common scenario, and we find that the inward-facing microphone receives a noticeable sound as soon as the heel touches the ground. So we need to filter out bone-conducted sounds of walking from the recorded sounds. Fig. 6 shows the PSD of the bone-conducted sound of walking, we can see that the frequency of the sound produced by heel landing is mainly concentrated below 100 Hz. The frequency difference between walking and occlusion is mainly caused by the different propagation paths. The sound produced by the heel landing needs to travel through the entire body to reach the ear canal, while the occlusal sound can reach the ear canal through a very short path. To remove the interference of walking, we adopt a band-pass filter ranging from 100 Hz to 2.5 kHz in *Air Noise Removal*. Thus, before the event detection, we can filter out bone-conducted sounds of walking.

### E. Feature Extraction

To accurately authenticate users and resist spoofing attacks, it is necessary to extract reliable biometric features from bone-conducted sounds of dental occlusion. In this section, we present approaches for extracting four features from the teeth structure, bone structure, occlusal location, and occlusal sound.

*Teeth Structure Feature:* When the upper and lower teeth collide with each other, surface acoustic waves (SAWs) are generated. SAWs can travel along the surface of a material, such as teeth and bone. Among them, the teeth structure features (e.g., enamel thickness, enamel rod patterns, and cusp morphology) have a direct impact on the acoustic signal caused by occlusion. According to Gnathosonics [58], different thicknesses of enamel can produce different frequencies when occlusion. Moreover, slight friction may occur between the teeth during occlusion, and the enamel rod patterns impact the friction and thus affect the sound of friction. The shape and quantity of cusps also contribute to the generation of occlusal sound.

To get effective biometric features of teeth structure, we extract the sonorant and fricative components of the occlusal sound. Specifically, the sonorant components are mainly influenced by enamel thickness and the cusp when performing occlusion, and the fricative components are mainly generated by enamel rod patterns on the tooth surfaces. To obtain the sonorant and fricative components, we calculate the cepstral coefficient of the occlusal sound. If the value is positive, the corresponding components in the sound have a high probability of sonorant components, otherwise, the corresponding components are contributed more by the fricative components. Fig. 7 shows the extracted sonorant and fricative components of two users.

*Bone Structure Feature:* Relevant research [59] shows that teeth and bone are dispersive mediums, which means that the propagation speed of SAWs is related to the frequency of SAWs and physical properties of the bone (e.g., density, elastic, and inertia properties). The speed of the high-frequency part of the SAWs is faster than that of the low-frequency part, which causes that the SAW spreads out and changes shape as it travels. Based on the above analysis, we try to extract the dispersion of occlusal sound as the structural feature related to physical properties of bones.

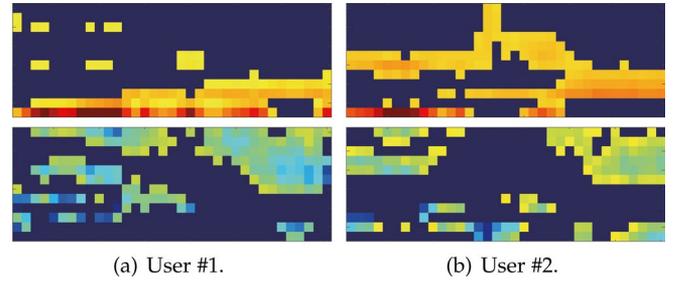

(a) User #1.   (b) User #2.

Fig. 7. Sonorant (upper) and fricative (lower) components of two users.

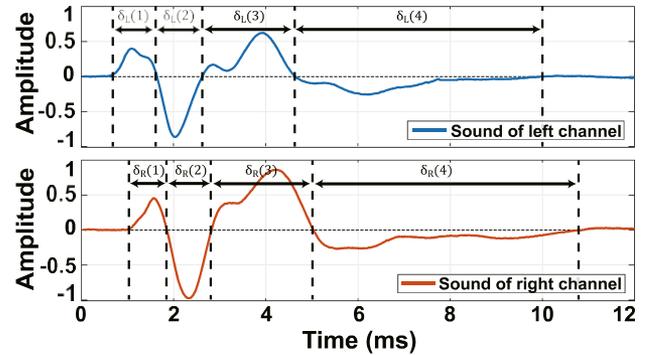

Fig. 8. Dispersion of two channels.

Fig. 8 shows the event of a dental occlusion detected by two inward-facing microphones in binaural canals. Due to dispersion, we can see that the high-frequency sound first reaches the microphones, then the low-frequency sound. Thus, we calculate the distance $\delta_L(i)$ (and $\delta_R(i)$) between two adjacent zero-crossing points. $ZS_L = [\delta_L(1), \delta_L(2), \ldots, \delta_L(n)]$ is defined as the zero-crossing sequence of left channel, then the zero-crossing sequence of right channel $ZS_R$ can be obtained. We regard $ZS_L$ and $ZS_R$ as the structural features related to the physical properties of bones.

*Location Feature:* When using the system, the user can choose any location of teeth to register, so the paths and times of occlusal sounds from different locations to the two microphones are also different. For instance, if the user occludes the left molars, the left microphone receives bone-conducted sound first. Since it is very difficult to accurately detect the number of teeth that complete the occlusion, we analyze the delay between the two occlusal sounds received by two microphones in both ears to indirectly get the location feature. However, different paths may cause different dispersion of occlusal sounds, which makes the delay calculation based on cross-correlation not accurate enough. To solve this problem, we first divide the occlusal sound into 5 frequency bands. For $i$-th frequency band, we compute cross-correlation $R_b(i)$ of the two occlusal sounds received by two microphones. Finally, we get the cross-correlation sequence $R_b = [R_b(1), R_b(2), \ldots, R_b(5)]$ as the location feature.

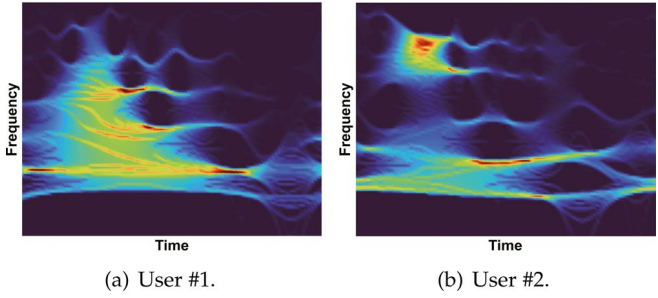

Fig. 9. SWT results of two users.

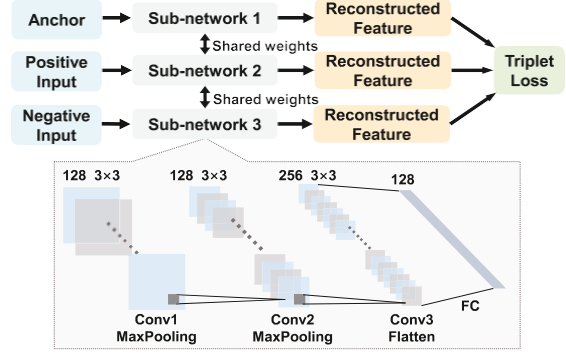

Fig. 10. Structure of the Triplet network.

*Acoustic Feature:* Finally, we extract the acoustic feature contained in the occlusal sound. At present, the most commonly used audio analysis methods include Mel-Frequency Cepstral Coefficients (MFCC) [46], Short-time Fourier transform (STFT) [47], and continuous wavelet transform (CWT) [48]. However, the window of STFT has a fixed length, which makes it unable to achieve the low-frequency and high-frequency features of the signal at the same time. Although CWT can adjust time resolution adaptively according to the frequency of the signal, the result of CWT is related to the selection of wavelet base. And the wavelet spectrum obtained by CWT has energy leakage, which makes the instantaneous frequency distribution fuzzy.

In order to obtain the acoustic feature contained in the occlusal sound, we study a time-frequency analysis method, Synchrosqueezing Wavelet Transform (SWT) [49]. Unlike other time-frequency analysis methods, SWT is based on wavelet transform and improves frequency resolution by compressing all frequency components, which preserves the time resolution and compensates for the spreading effects caused by the wavelet base. To calculate SWT, we first use CWT to obtain the wavelet coefficient $W(a, b)$ ($a$ is the scale factor, $b$ is the time translation) for the input sound signal $x(t)$. Then we extract the instantaneous frequency by using the phase transformation, which is defined as

$$\omega(a, b) = -\mathrm{i} W(a, b)^{-1} \frac{\partial W(a,b)}{\partial b}, \quad (4)$$

where $\omega(a, b)$ denotes the instantaneous frequency, and i is the imaginary unit. Then, SWT transfers the information from the time-scale plane to the time-frequency plane. Suppose $\omega_l$ is the closest frequency to the original frequency $\omega(a, b)$, each value of $W(a, b)$ is reallocated into $T(\omega_l, b)$ as

$$T(\omega_l, b) = \frac{1}{\Delta\omega} \sum_{a_k : |\omega(a_k, b) - \omega_l| < \frac{\Delta\omega}{2}} W(a_k, b) a_k^{-\frac{3}{2}} (\Delta a)_k, \quad (5)$$

where $\Delta\omega$ denotes the width of frequency bins ($\omega_l - \Delta\omega/2, \omega_l + \Delta\omega/2$). $\omega_l$ is the discrete frequency variable and $\Delta\omega = \omega_l - \omega_{l-1}$. $a_k$ is the discrete wavelet scale and $(\Delta a)_k = a_k - a_{k-1}$. Fig. 9 shows the results calculated by SWT of two users. Then, the results are converted into grayscale images. Due to the use of two microphones, we can get two grayscale images. Finally, we combine the two images into a two-channel image as the acoustic feature.

### F. Authentication

*Authentication Model:* After getting four biometric features, we design a two-stage approach for user authentication. First, we use teeth structure features and acoustic features to complete the first step of authentication. Traditional classifiers for authentication (e.g., SVM, RF, and DNN) usually have a fixed number of output classes and need a large number of positive and negative data for training. So they are not suitable for solving the problem that the number of output classes is uncertain and the training data set is small, such as face recognition and user authentication. To solve this problem, we leverage a Triplet network [25] as the classification model to classify the teeth structure features and acoustic features. The Triplet network includes three identical sub-networks. The input is a triplet consisting of an anchor, a positive input, and a negative input. Specifically, the features of the same user are the anchor and the positive input, and the feature of another user is the negative input. The output of each sub-network is a reconstructed feature vector that can represent user identity. The training goal is to minimize the distance between two reconstructed feature corresponding to the same user, and maximize the distance between two reconstructed feature corresponding to different users.

Fig. 10 shows the structure of the Triplet network, which has three identical and weight-sharing sub-networks. Each sub-network mainly consists of three convolution (Conv) layers, two maxpooling layers, a flatten layer, and a fully connected (FC) layers. The Conv layer is used to extract deep features from the input. A maxpooling layer is added between two Conv layers to reduce feature size. Given three features as the inputs, the Triplet network can extract reconstructed features related to user identities through three sub-networks, and computes the triplet loss as the similarity of the inputs. The triplet loss of the network can be expressed as

$$L = \max(d(F_a, F_p) - d(F_a, F_n) + \upsilon, 0), \quad (6)$$

where $d(F_a, F_p)$ denotes the euclidean distance between two reconstructed features from sub-network 1 and sub-network 2,

$d(F_a, F_n)$ denotes the euclidean distance between two reconstructed feature from sub-network 1 and sub-network 3, and $v$ is a threshold that represents the minimum distance. By minimizing the triplet loss, we can minimize the distance between the features of the same user and maximize the distance between the features of different users.

Before our system is deployed, we first pre-train the network to make it have basic authentication ability. We ask 4 volunteers to find comfortable occlusal locations and practice occlusion a few times. Then we use a pair of inward-facing microphones to collect bone-conducted sounds of occlusion. After extracting features, we construct triplets and input them into the Triplet network for pre-training. After that, the network has the preliminary ability to distinguish teeth structure features and acoustic features from different users.

*User Registration:* When new users register, they need to perform occlusion several times to extract the teeth structure features and acoustic features. To reduce the time of occlusion, we augment the features by using frequency-time and time-warping masking [28], [61]. Then, the features of each registrant, the 4 volunteers, and other registered users (if any) are used to form new triplets, which are then send to the network to retrain it. To reduce training cost, we borrow ideas from incremental learning [26]. Specifically, we only use the new triplets input to fine tune the network parameters, rather than completely retrain all parameters of the network. Since the three sub-networks have the same parameters, we only need to use one trained sub-network to obtain the reconstructed features after the training. We then calculate an average reconstructed feature for each registrant, which is defined as the average of reconstructed features extracted from the registrant's registration data. In the registration phase, we also collect bone structure features of each registrant and average zero-crossing sequences to get $\tilde{Z}S_L$ and $\tilde{Z}S_R$. For the location feature, we calculate average cross-correlation sequence $\tilde{R}_b$. Finally, the average reconstructed feature, average zero-crossing sequences, and average cross-correlation sequence are stored in the user template.

*User Login:* In the login phase, the sub-network first extracts the reconstructed feature of the loginer, we then compare the euclidean distances between the reconstructed feature and each average reconstructed feature in the user template. If the distances are all larger than a threshold, we consider there is a spoofer, otherwise, the identity is assigned to the registered user with the the minimum distance. Next, we calculate zero-crossing difference between $ZS_L$ of the received bone structure feature and $\tilde{Z}S_L$ of the registered user assigned by the network as $DS_L = \frac{1}{n}\sum_{k=1}^{n}[\delta_L(k) - \tilde{\delta}_L(k)]$. Then we get $DS_R$ of the right channel in the same way. For the location feature, we use the same method to calculate cross-correlation difference $DR_{L,R}$. If any two of $DS_L$, $DS_R$, and $DR_{L,R}$ are less than corresponding thresholds, the user is considered legitimate.

## V. IMPLEMENTATION AND EVALUATION

In this section, we introduce the implementation details and provide the evaluation results.

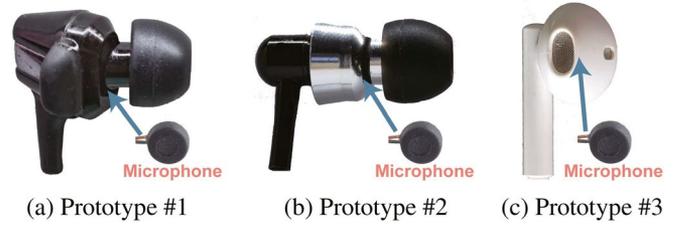

(a) Prototype #1  (b) Prototype #2  (c) Prototype #3

Fig. 11. Prototype earphones.

### A. Experiment Setup

There are several commercial earbuds equipped with inward-facing microphones. But the hardware manufacturers do not open the corresponding permissions for us to get sound data from inward-facing microphones. Thus, we implement TeethPass+ by attaching a microphone in front of the speaker in an earphone, which is similar to most commercial earbuds. We design 3 prototypes, as shown in Fig. 11. We recruit 53 volunteers (34 males and 19 females, aged from 18 to 61), with 45 of them registering TeethPass+ as legitimate users, and the rest 8 volunteers are spoofers. Each legitimate user chooses a comfortable prototype earphone and occlusal location. The occlusal sounds are recorded in 4 environments, including lab, park, car, and mall. Finally, we collect more than 5,000 occlusal sounds for legitimate users. Part of the data collected in the lab are used to train the network, and the remaining data are used as the test set. Each spoofer performs 100 times of mimic attacks, replay attacks, and hybrid attacks to all legitimate users. All procedures are approved by the Institutional Review Board (IRB) at our institute.

### B. Evaluation Methodology

We mainly evaluate TeethPass+ from the following aspects.
*Confusion Matrix:* Each row and each column of the matrix represent the ground truth and the authentication result, respectively. Each entry represents the percentage of a user that is classified into each identity.

*False Reject Rate (FRR):* The probability that TeethPass+ authenticates a legitimate user as a spoofer.

*False Accept Rate (FAR):* The probability that TeethPass+ authenticates a spoofer as a legitimate user.

### C. Overall Performance

We first evaluate the overall performance of TeethPass+. Fig. 12 shows the confusion matrix of 45 legitimate users (denoted as $U_1, U_2, \ldots, U_{45}$) and 8 spoofers (denoted as $SP$) in all environments. We sort 45 legitimate users. $U_1$ to $U_{18}$ use prototype #1, $U_{19}$ to $U_{33}$ use prototype #2, $U_{34}$ to $U_{45}$ use prototype #3. It can be seen that TeethPass+ achieves an average accuracy of 98.6% for user authentication and 99.7% for the detection of 3 types of attacks. Among the 45 legitimate users, the lowest accuracy is 95.2% (user #39), which is acceptable for an authentication system. The results indicate that TeethPass+

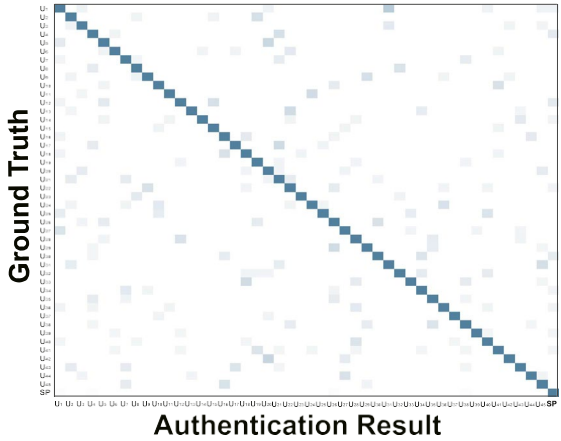

Fig. 12. Confusion matrix of TeethPass+.

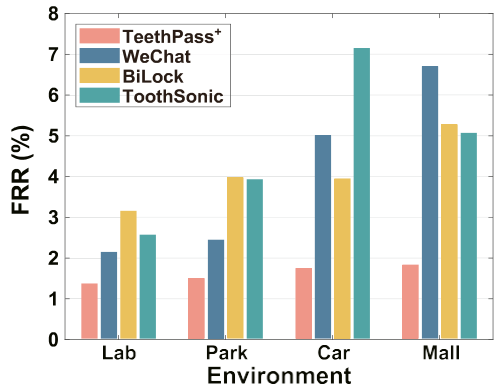

Fig. 13. FRR of TeethPass+ and comparison methods under 4 environments.

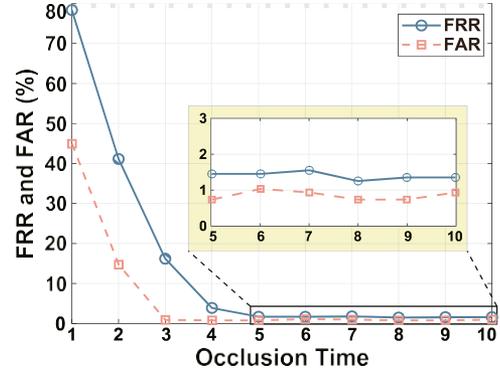

Fig. 14. FRR and FAR under different occlusion times for registration.

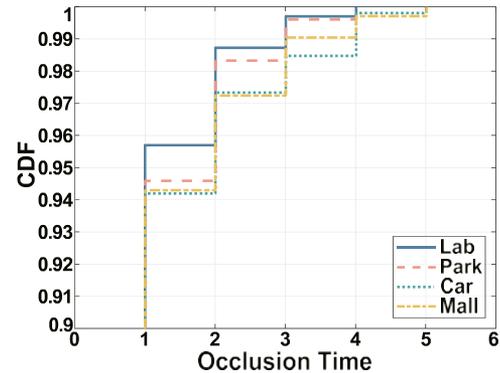

Fig. 15. CDF of occlusion times for login.

can accurately authenticate legitimate users and detect spoofers in various environments.

We compare the performance of TeethPass+ with that of WeChat voiceprint lock [62], BiLock [33], and ToothSonic [32]. Fig. 13 shows the FRR of the 4 systems in 4 environments, respectively. We can see that the FRR of TeethPass+ is 1.3%, which is slightly better than 2.1%, 3.1%, and 2.5% of three comparison methods in the lab. But in cars and malls with loud ambient noises, the FRR of comparison methods increase over 3.9%, while TeethPass+ keeps a stable FRR with a slight increase to 1.7%. With the help of noise reduction methods, TeethPass+ has stable performance in various environments. Another reason is that the shell of the earphone also helps to isolate ambient noises. Moreover, since WeChat voiceprint lock and BiLock are all based on air-conducted sounds, their anti-replay-attack ability is significantly weaker than our system.

### D. Performance on User Experience

*Occlusion Time for Registration:* In the registration phase, more times of occlusion can improve the effect of training the network. But too many times of occlusion may lead to a poor user experience. Hence, we evaluate the FRR and FAR of TeethPass+ under different occlusion times for registration, and the results are shown in Fig. 14. It is obvious that with the occlusion time for registration increases, the FRR and FAR of TeethPass+ decrease sharply at the beginning. And when the time reaches 5 and 3, the FRR and FAR do not decrease with the increase of occlusion times, respectively. TeethPass+ only needs 5 occlusion to achieve 1.4% FRR and 0.7% FAR, which is mainly because we design a suitable Triplet network and adopt data augmentation. In order to balanced trade-off between performance and usability, we fix the occlusion times to 5 in all the evaluations.

*Occlusion Time for Successful Login:* In the login phase, we evaluate the occlusion times required for successful authentication under 4 environments. Fig. 15 shows CDF of the occlusion times. We can see that more than 94.1% of login operations are successful with only one occlusion under each environment. And an average of 99.8% of login operations can be successfully authenticated within 4 occlusion, which is acceptable for users. In some special environments, such as a very loud sound source near the user, TeethPass+ requires 5 or more occlusion to successfully authenticate the user. Therefore, for the sake of security, when there occurs 4 consecutive times of unsuccessful authentication, the device is automatically locked for a while.

*Authentication Response Time:* Then we study the response time of TeethPass+ from receiving an occlusal sound to producing the authentication result. During the evaluation, all the registration and login data are transmitted to a PC with a 3.2 GHz

TABLE I
RESPONSE TIME DURING AN AUTHENTICATION

| Phase | ANR | ED | MR | FE | Train | Auth | Total |
|---|---|---|---|---|---|---|---|
| Reg(s) | 0.3 | 0.2 | 0.1 | 0.3 | 3.6 | None | 4.5 |
| Login(ms) | 58 | 45 | 28 | 79 | None | 63 | 273 |

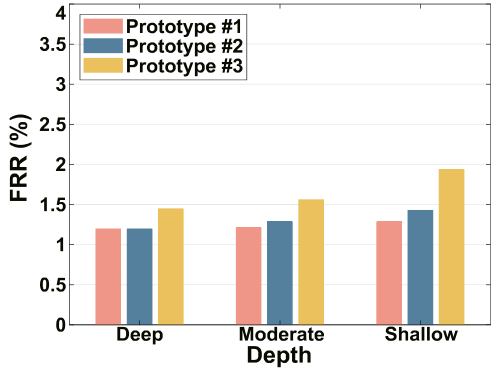

Fig. 16. FRR under different wearing depths.

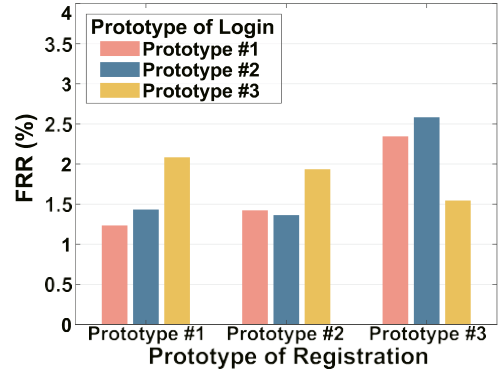

Fig. 17. FRR under different prototypes.

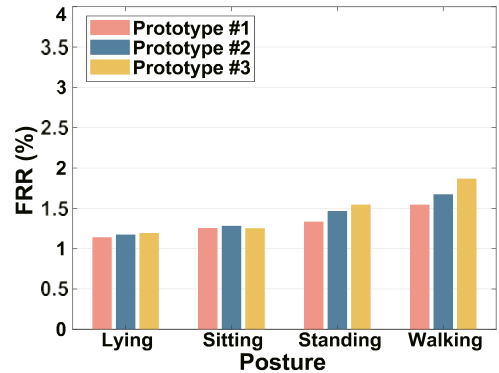

Fig. 18. FRR under different user postures.

Intel i7 CPU and 16 GB memory. Table I shows the average response times of *Air Noise Removal* (ANR), *Event Detection* (ED), *Motion Removal* (MR), *Feature Extraction* (FE), *Network Training* (Train), and *Authentication* (Auth), respectively. It can be seen that in the registration (Reg) phase, the training of the Triplet network takes the most time, and the total response time for each new user to register TeethPass+ is about 4.5 s. But response time of registration has little impact on user experience. In the login phase, TeethPass+ can produce authentication results within 273 ms after the user completes the occlusion, which indicates that TeethPass+ can achieve a satisfactory user experience.

*Earphone Wearing Depth:* The types and depths of earphones that each user is accustomed to wearing are different, so we study the impact of different earphone types and depths on system performance. Each user chooses a comfortable earphone from the 3 prototypes. The occlusal sounds are collected at different in-ear depths, including deep position, moderate position, and shallow position. It can be seen from Fig. 16 that prototype #1 has the lowest FRR at all three depths since it has the most stable wearing way. The FRR of prototype #2 increases slightly at shallow depth, but its in-ear structure can ensure that the FRR is less than 1.5%. Prototype #3 adopts a semi-in-ear structure, so its wearing stability and sound insulation are not good, causing FRR to rise to about 1.9% at shallow depth. In general, the type of earphone has a certain impact, while the wearing depth of the earphone has little effect on performance.

*Earphone Structure:* To evaluate the system performance when users register and log in with different earphones, we require volunteers to register using one prototype and then log in using different prototypes. The results are shown in Fig. 17. We can see that if the volunteers use the same prototype to register and log in, the worst FRR is about 1.5% (prototype #3). But when volunteers log in using a prototype different from the registered one, there is a significant increase in FRR. Especially when using prototype #3 for registration, the system performance decreases when using prototypes #1 and #2 for login, which may be due to the significant difference in the structure of prototype #3 compared to #1 and #2. To identify the specific earphones in use, we obtain the unique Bluetooth address in the Bluetooth earphone or the device ID in the USB earphone. We require the earphones used at the time of registration to match the earphones used during login.

*User Posture:* Next, we evaluate the system performance under different user postures, including lying, sitting, standing, and walking. The authentication results are shown in Fig. 18. We can see that when users maintain static postures (i.e., lying, sitting, and standing), the FRR of the three prototypes is within 1.5%. But if users are in a dynamic posture (i.e., walking), the FRR of three prototypes all increase. Although we remove the bone-conducted sound of walking in Section IV-D, some body vibrations can also transmit to the earphone. Even so, the FRR increases to about 1.8% at most (Prototype #3).

*Occlusal Force:* Although users are required to use consistent force to complete the occlusion during the registration and login phase, we still conduct additional experiments to verify the robustness of our system when using different forces during registration and login. We divide the force of occlusion into three levels: light, moderate, and hard. Different levels of force are used to form registration data and login data, respectively. The authentication results are shown in Fig. 19. After analysis, we find that when using light force for registration, hard occlusion during login may lead to higher FRR. And when using hard force for registration, light occlusion during login also leads to higher FRR. However, if the user uses moderate force to complete the registration, the system performance is not seriously reduced by using light or hard force during login. Therefore, we recommend that users use moderate force to complete the occlusion when registering our system.

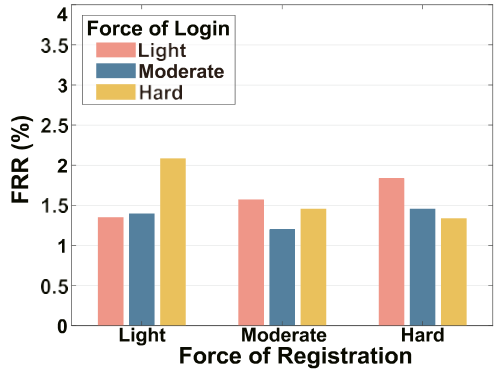

Fig. 19. FRR under different occlusal force.

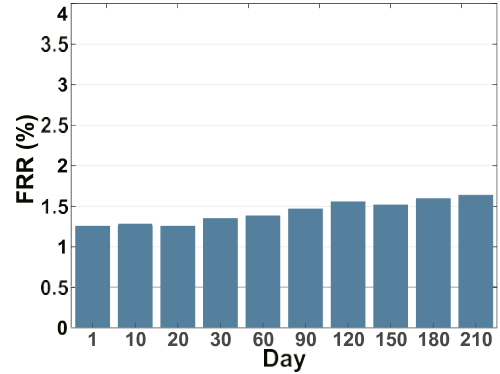

Fig. 21. FRR over different time of period.

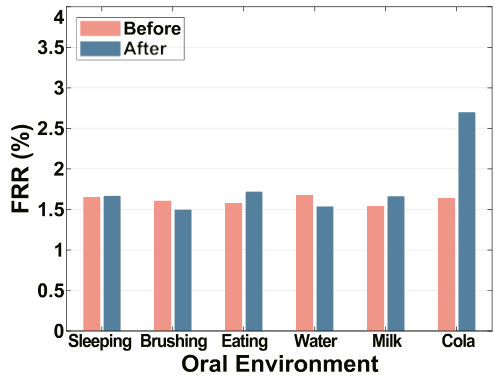

Fig. 20. FRR under different oral environments.

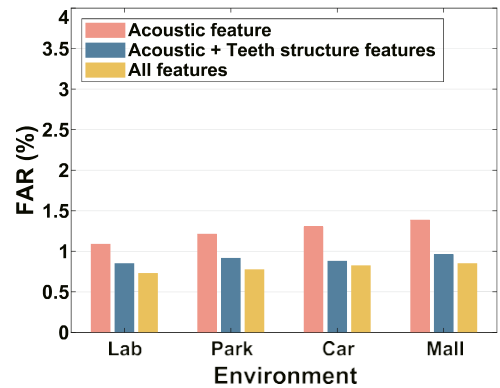

Fig. 22. FAR under mimic attacks.

*Oral Environment:* We also evaluate the impact of different oral environments on system performance. We ask users to use TeethPass+ before and after sleeping, brushing teeth, eating, drinking water, drinking milk, and drinking cola, respectively. Fig. 20 shows that the FRR after sleeping, eating, and drinking milk increases slightly, while brushing teeth and drinking water can reduce the FRR. In particular, we find that drinking cola has a great impact on system performance, the FRR after drinking cola increases to 2.7%. The reason may be that the ingredients with high viscosity (e.g., caramel color and syrup) in cola adhere to the surface of teeth, resulting in the change of biometric features of the occlusion.

In addition, the oral environment may change over time. To evaluate the robustness of TeethPass+ over time, we collect occlusal sounds of the users for more than 7 months. Fig. 21 shows the authentication FRR of different days after registration. We can observe that TeethPass+ can maintain high performance over a long time. Specifically, the FRR of less than 1.5% after 90 days. Even over 7 months, the system performance can maintain a very low FRR less than 1.6%. Moreover, it is worth noticing that two users extract a wisdom teeth during the experiment, and the FRR after teeth extraction does not increase significantly.

### E. Performance on Attack Resistance

In order to prove that TeethPass+ can resist attacks described in Section III-C, we conduct several experiments to verify the effectiveness of the three biometric features under the mimic attack, replay attack, and hybrid attack, respectively. Spoofers use the same prototype earphone as the legitimate user to attack TeethPass+.

*Mimic Attack:* To conduct mimic attacks, we assume that spoofers know which teeth and how much force the legitimate user uses to occlude. Fig. 22 shows the FAR of mimic attacks under different environments. It shows that the FAR is less than 1.1% in the lab when using only acoustic feature. But when the acoustic feature and teeth structure feature are used, the average FAR is stable at about 0.8%. And the FAR is further reduced when using all three features, which indicates that TeethPass+ using all three features can resist mimic attacks well. The reason is that although the spoofer can mimic the location and force of occlusion of the legitimate user, the biometric features of the spoofer's teeth and bones are still different from those of the legitimate user, resulting in acoustic and structure features that are not identical to legitimate users.

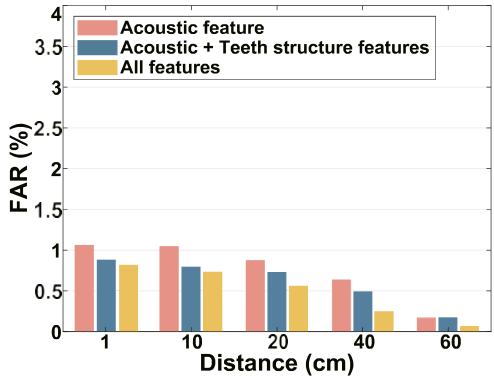

Fig. 23. FAR under replay attacks.

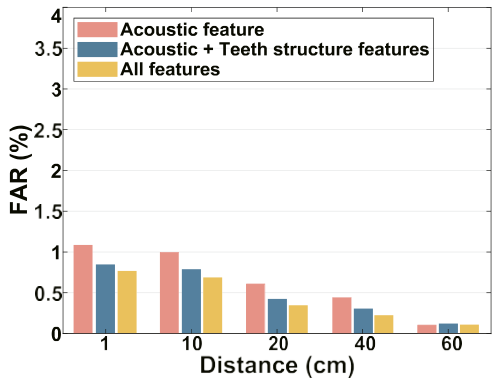

Fig. 24. FAR under hybrid attacks.

*Replay Attack:* To conduct replay attacks, we place a microphone at different distances to the user's mouth to eavesdrop on the air-conducted sound of dental occlusion and then replay it to the prototype earphone to attack. Fig. 23 depicts the results of replay attacks at different eavesdropping distances. It is obvious that even if the eavesdropping microphone is very close to the mouth (about 1 cm), the FAR does not exceed 0.8% by using all features. When the distance is greater than 20 cm, the FAR is reduced to about 0.5% by using all features. Since the air-conducted sounds of occlusion are more close to impulse waves and decay fast, the FAR decreases rapidly with the increase of distance. In addition, due to the fact that the air-conducted sound of occlusion has completely different biometric features than bone-conducted sound of occlusion, the FAR of replay attack is lower than that of mimic attack.

*Hybrid Attack:* Finally, we consider that the spoofer mimics the occlusion of the user while playing the eavesdropped occlusal sound by a bone conduction speaker in the spoofer's mouth. The setting of eavesdropping is the same as that in replay attacks. Fig. 24 shows the results of hybrid attacks at different eavesdropping distances. We can see that when the distance is greater than 20 *cm*, the FAR is reduced to around 0.5%. At smaller distances, the FAR of hybrid attacks is less than 1%. But in actual scenarios, it is difficult to eavesdrop on users in such a short distance. Generally, TeethPass+ can resist various attacks effectively in different environments and eavesdropping distances.

## VI. Discussion and Limitations

We evaluate several factors that affect system performance, but there are also several limitations and we discuss the opportunities to overcome them.

*Single Earphone Authentication:* In order to extract the three types of biometric features, we require users to wear two earphones at the same time. However, after investigation, we find that some users only wear single earphone in some cases, which makes our system unable to extract all the features. In this case, we can only extract the structural feature and acoustic feature of one channel. By analyzing the attack resistance experiments (in Section V-E), we can find that even if only part of features are adopted, our system can still effectively resist most of the attacks. In future, we will use the registration data of single channel to train user template of single channel, and decide whether to use user template of two channel or single channel according to the wearing condition.

*Incompatible With Speakers:* The diaphragm of the speaker vibrates when the earphone plays sound and the vibration can spread to the inward-facing microphone, thus the sound played by the speaker affects the inward-facing microphone to sense the bone-conducted sounds of occlusion. In this paper, all of experiments are conducted without other sounds being played by the earphones. As our future work, we will try to remove the interference of the sound played by the earphones by analyzing the correlation between the played and received sounds. Moreover, combining with other authentication methods may help to further improve the usability of our system.

*Long-Term Experiments:* Our experiments lasting more than 7 months demonstrate the effectiveness of our system, but the biometric features of teeth may change after several years, which adversely affect performance of our system. To solve this problem, we can continue to periodically collect biometric features of registered users, and retrain the triplet network through incremental learning to update the user template stored in our system.

*Usability Study:* Finally, we ask the volunteers to answer a questionnaire. The questions are based on System Usability Scale (SUS) and modified according to our system. The questions are about user comfort of wearing the prototypes and performing the occlusion as well as the usability in society. Each volunteer gives a score between 1 to 5 (with 5 being the highest) to each of the following items: I will use TeethPass+ frequently (Q1), TeethPass+ is easy to use (Q2), it is comfortable to perform occlusion (Q3), TeethPass+ does not affect the normal use of earphones (Q4). As shown in Table II, the result shows that the users strongly approve of our prototype and idea. All users agree that TeethPass+ is easy to use and will be used frequently. Most users can find at least one comfortable occlusal location. We also find that a small number of users are not satisfied that they cannot play music and authenticate at the same time. However, TeethPass+ can combine with other authentication methods to further improve usability, and most users can still accept it. For example, Earmonitor [63] can perform user identification based on the subtle changes of the ear canal even when the user is listening to music. However, the method requires earphones to emit ultrasound, which may impair the health of users experi- encing long-term exposure. Thus, TeethPass+ is better-suited for long-term us than Earmonitor.

TABLE II
USABILITY SCORES

|    | Mean | Standard Deviation | Min  | Max  |
|----|------|--------------------|------|------|
| Q1 | 4.15 | 0.60               | 3.00 | 5.00 |
| Q2 | 4.31 | 0.46               | 4.00 | 5.00 |
| Q3 | 3.97 | 0.69               | 2.00 | 5.00 |
| Q4 | 3.68 | 0.73               | 2.00 | 5.00 |

## VII. CONCLUSION

In this paper, we design and implement TeethPass+, which uses inward-facing microphones in earbuds to collect bone-conducted sounds of dental occlusion in binaural canals to achieve authentication. We present effective methods to filter out interferences of ambient noises and daily actions. We extract biometric features, including physical features of teeth and bone, location features of occlusion, and acoustic features of occlusal sound, then adopt a Triplet network based on incremental learning as the authentication model. The extensive experiments show that it achieves an average authentication accuracy of 98.6%, and resists 99.7% of spoofing attacks.

## REFERENCES

[1] Cisco, "2020 Cisco consumer privacy survey," 2020. [Online]. Available: https://www.freevacy.com/news/cisco/2020-cisco-consumer-privacy-survey/894
[2] IBM, "2020 Cost of a data breach report," 2020. [Online]. Available: https://securityintelligence.com/posts/whats-new-2020-cost-of-a-data-breach-report/
[3] J. Yu, L. Lu, Y. Chen, Y. Zhu, and L. Kong, "An indirect eavesdropping attack of keystrokes on touch screen through acoustic sensing," *IEEE Trans. Mobile Comput.*, vol. 20, no. 2, pp. 337–351, Feb. 2021.
[4] N. K. Ratha, V. D. Pandit, R. M. Bolle, and V. Vaish, "Robust fingerprint authentication using local structural similarity," in *Proc. IEEE 5th Workshop Appl. Comput. Vis.*, 2000, pp. 29–34.
[5] K. K. M. Shreyas, S. Rajeev, K. Panetta, and S. S. Agaian, "Fingerprint authentication using geometric features," in *Proc. IEEE Int. Symp. Technol. Homeland Secur.*, 2017, pp. 1–7.
[6] G. Heigold, I. Moreno, S. Bengio, and N. Shazeer, "End-to-end text-dependent speaker verification," in *Proc. IEEE Int. Conf. Acoust. Speech Signal Process.*, 2016, pp. 5115–5119.
[7] R. G. M. M. Jayamaha, M. R. R. Senadheera, T. N. C. Gamage, K. D. P. B. Weerasekara, G. A. Dissanayaka, and G. N. Kodagoda, "VoizLock – Human voice authentication system using hidden Markov model," in *Proc. IEEE 4th Int. Conf. Inf. Automat. Sustainability*, 2008, pp. 330–335.
[8] P. L. D. Leon, M. Pucher, J. Yamagishi, I. Hernaez, and I. Saratxaga, "Evaluation of speaker verification security and detection of HMM-based synthetic speech," *IEEE Trans. Audio, Speech, Lang. Process.*, vol. 20, no. 8, pp. 2280–2290, Oct. 2012.
[9] I. Song, H. J. Kim, and P. B. Jeon, "Deep learning for real-time robust facial expression recognition on a smartphone," in *Proc. IEEE Int. Conf. Consum. Electron.*, 2014, pp. 564–567.
[10] I. Masi, Y. Wu, T. Hassner, and P. Natarajan, "Deep face recognition: A survey," in *Proc. 31st SIBGRAPI Conf. Graph. Patterns Images*, 2018, pp. 471–478.
[11] Apple, "Use touch ID on iPhone and iPad," 2020. [Online]. Available: https://support.apple.com/en-us/HT201371
[12] TD Bank, "TD voicePrint," 2021. [Online]. Available: https://www.td.com/privacy-and-security/privacy-and-security/how-we-protect-you/weprotect.jsp
[13] Amazon, "Amazon Rekognition," 2017. [Online]. Available: https://aws.amazon.com/cn/rekognition/
[14] H. Kong, L. Lu, J. Yu, Y. Chen, and F. Tang, "Continuous authentication through finger gesture interaction for smart homes using WiFi," *IEEE Trans. Mobile Comput.*, vol. 20, no. 11, pp. 3184–3162, Nov. 2021.
[15] M. Zhou et al., "PressPIN: Enabling secure PIN authentication on mobile devices via structure-borne sounds," *IEEE Trans. Dependable Secure Comput.*, vol. 20, no. 2, pp. 1228–1242, Mar./Apr. 2023.
[16] C. Wu, K. He, J. Chen, Z. Zhao, and R. Du, "Toward robust detection of puppet attacks via characterizing fingertip-touch behaviors," *IEEE Trans. Dependable Secure Comput.*, vol. 19, no. 6, pp. 4002–4018, Nov./Dec. 2022.
[17] H. Jiang, H. Cao, D. Liu, J. Xiong, and Z. Cao, "SmileAuth: Using dental edge biometrics for user authentication on smartphones," in *Proc. ACM Interactive Mobile Wearable Ubiquitous Technol.*, vol. 4, no. 3, pp. 84:1–84:24, 2020.
[18] L. Lu et al., "LipPass: Lip reading-based user authentication on smartphones leveraging acoustic signals," in *Proc. IEEE Conf. Comput. Commun.*, 2018, pp. 1466–1474.
[19] Y. Gao, W. Wang, V. V. Phoha, W. Sun, and Z. Jin, "EarEcho: Using ear canal echo for wearable authentication," in *Proc. ACM Interactive Mobile Wearable Ubiquitous Technol.*, vol. 3, no. 3, pp. 81:1–81:24, 2019.
[20] M. Li, I. Cohen, and S. Mousazadeh, "Multisensory speech enhancement in noisy environments using bone-conducted and air-conducted microphones," in *Proc. IEEE China Summit Int. Conf. Signal Inf. Process.*, 2014, pp. 1–5.
[21] B. Huang, Y. Gong, J. Sun, and Y. Shen, "A wearable bone-conducted speech enhancement system for strong background noises," in *Proc. IEEE 18th Int. Conf. Electron. Packag. Technol.*, 2017, pp. 1682–1684.
[22] A. Shahina and B. Yegnanarayana, "Language identification in noisy environments using throat microphone signals," in *Proc. Int. Conf. Intell. Sens. Inf. Process.*, 2005, pp. 400–403.
[23] Canalys, "Global smart accessories forecast 2021," 2021. [Online]. Available: https://www.canalys.com/newsroom/global-smart-accessories-market-2021-forecast
[24] H. Kim, A. Byanjankar, Y. Liu, Y. Shu, and I. Shin, "UbiTap: Leveraging acoustic dispersion for ubiquitous touch interface on solid surfaces," in *Proc. 16th ACM Conf. Embedded Netw. Sensor Syst.*, 2018, pp. 211–223.
[25] E. Hoffer and N. Ailon, "Deep metric learning using triplet network," in *Proc. Int. Workshop Similarity-Based Pattern Recognit.*, 2015, pp. 84–92.
[26] Y. Xie, F. Li, Y. Wu, and Y. Wang, "HearFit: Fitness monitoring on smart speakers via active acoustic sensing," in *Proc. IEEE Conf. Comput. Commun.*, 2021, pp. 1–10.
[27] P. Jiang et al., "Securing liveness detection for voice authentication via pop noises," *IEEE Trans. Dependable Secure Comput.*, vol. 20, no. 2, pp. 1702–1718, Mar./Apr. 2023.
[28] Y. Gao, Y. Jin, J. Chauhan, S. Choi, J. Li, and Z. Jin, "Voice in ear: Spoofing-resistant and passphrase-independent body sound authentication," in *Proc. ACM Interactive Mobile Wearable Ubiquitous Technol.*, vol. 5, no. 1, pp. 12:1–12:25, 2021.
[29] D. Liu et al., "SoundID: Securing mobile two-factor authentication via acoustic signals," *IEEE Trans. Dependable Secure Comput.*, vol. 20, no. 2, pp. 1687–1701, Mar./Apr. 2023.
[30] M. P. G. Salazara and J. R. Gasgaa, "Microhardness and chemical compo- sition of human tooth," *Mater. Res.*, vol. 6, no. 3, pp. 367–373, 2003.
[31] D. S. Kim, K. W. Chung, and K. S. Hong, "Person authentication using face, teeth and voice modalities for mobile device security," *IEEE Trans. Consum. Electron.*, vol. 56, no. 4, pp. 2678–2685, Nov. 2010.
[32] Z. Wang, Y. Ren, Y. Chen, and J. Yang, 'ToothSonic: Earable authentication via acoustic toothprint," in *Proc. ACM Interactive Mobile Wearable Ubiquitous Technol.*, vol. 6, no. 2, pp. 78:1–78:24, 2022.
[33] Y. Zou, M. Zhao, Z. Zhou, J. Lin, M. Li, and K. Wu, "BiLock: User authentication via dental occlusion biometrics," in *Proc. ACM Interactive Mobile Wearable Ubiquitous Technol.*, vol. 2, no. 3, pp. 152:1–152:20, 2018.
[34] T. Arakawa, T. Koshinaka, S. Yano, H. Irisawa, R. Miyahara, and H. Imaoka, "'Fast and accurate personal authentication using ear acoustics," in *Proc. Asia-Pacific Signal Inf. Process. Assoc. Annu. Summit Conf.*, 2016, pp. 1–4.
[35] Z. Wang, S. Tan, L. Zhang, Y. Ren, Z. Wang, and J. Yang, "EarDynamic: An ear canal deformation based continuous user authentication using in-ear wearables," in *Proc. ACM Interactive Mobile Wearable Ubiquitous*

[36] T. G. Leighton, "Are some people suffering as a result of increasing mass exposure of the public to ultrasound in air?," *Roy. Soc. A Math. Phys. Eng. Sci.*, vol. 472, no. 2185, pp. 1–57, 2016.
[37] Y. Xie, F. Li, Y. Wu, H. Chen, Z. Zhao, and Y. Wang, "TeethPass: Dental occlusion-based user authentication via in-ear acoustic sensing," in *Proc. IEEE Conf. Comput. Commun.*, 2022, pp. 1789–1798.
[38] K. Krishan, T. Kanchan, and A. K. Garg, "Dental evidence in forensic identification – An overview, methodology and present status," *Open Dent. J.*, vol. 9, no. 1, pp. 250–256, 2015.
[39] S. Hillson and S. Bond, "Relationship of enamel hypoplasia to the pattern of tooth crown growth: A discussion," *Amer. J. Phys. Anthropol.*, vol. 104, no. 1, pp. 89–103, 1997.
[40] T. M. Smith, A. J. Olejniczak, D. J. Reid, R. J. Ferrell, and J. J. Hublin, "Modern human molar enamel thickness and enamel–dentine junction shape," *Arch. Oral Biol.*, vol. 51, no. 11, pp. 974–995, 2006.
[41] J. Naziya, S. Sunil, P. Jayanthi, R. Rathy, and R. K. Harish, "Analysis of enamel rod end pattern for personal identification," *J. Oral Maxillofac. Pathol.*, vol. 23, no. 1, pp. 1–5, 2019.
[42] A. Ortiz, S. E. Bailey, G. T. Schwartz, J. Hublin, and M. M. Skinner, "Evo-devo models of tooth development and the origin of hominoid molar diversity," *Sci. Adv.*, vol. 4, no. 4, pp. 1–6, 2018.
[43] S. Davies and R. M. J. Gray, "What is occlusion?," *Brit. Dent. J.*, vol. 191, no. 5, pp. 235–245, 2001.
[44] W. E. Siri, "The gross composition of the body," *Elsevier Adv. Biol. Med. Phys.*, vol. 4, pp. 239–280, 1956.
[45] Y. Xie, F. Li, Y. Wu, and Y. Wang, "HearFit$^+$: Personalized fitness monitoring via audio signals on smart speakers," *IEEE Trans. Mobile Comput.*, vol. 22, no. 5, pp. 2756–2770, May 2023.
[46] Y. Wu, F. Li, Y. Xie, S. Yang, and Y. Wang, "HDSpeed: Hybrid detection of vehicle speed via acoustic sensing on smartphones," *IEEE Trans. Mobile Comput.*, vol. 21, no. 8, pp. 2833–2846, Aug. 2022.
[47] D. Gabor, "Theory of communication. Part 1: The analysis of information," *J. Inst. Elect. Engineers-Part III: Radio Commun. Eng.*, vol. 93, no. 26, pp. 429–441, 1946.
[48] I. Daubechies, "The wavelet transform, time-frequency localization and signal analysis," *IEEE Trans. Inf. Theory*, vol. 36, no. 5, pp. 961–1005, Sep. 1990.
[49] I. Daubechies, J. Lu, and H.-T. Wu, "Synchrosqueezed wavelet transforms: An empirical mode decomposition-like tool," *Elsevier Appl. Comput. Harmon. Anal.*, vol. 30, no. 2, pp. 243–261, 2011.
[50] M. Chen et al., "Your table can be an input panel: Acoustic-based device-free interaction recognition," in *Proc. ACM Interactive Mobile Wearable Ubiquitous Technol.*, vol. 3, no. 1, pp. 3:1–3:21, 2019.
[51] X. Xu et al., "TouchPass: Towards behavior-irrelevant on-touch user authentication on smartphones leveraging vibrations," in *Proc. 26th Annu. Int. Conf. Mobile Comput. Netw.*, 2020, pp. 1–13.
[52] Advanced Television Systems Committee, "Techniques for establishing and maintaining audio loudness for digital television," 2013. [Online]. Available: https://www.atsc.org/atsc-documents/type/1--0-standards/
[53] M. O. Culjat, D. Goldenberg, P. Tewari, and R. S. Singh, "A review of tissue substitutes for ultrasound imaging," *Ultrasound Med. Biol.*, vol. 36, no. 6, pp. 861–843, 2010.
[54] M. Jalil, F. A. Butt, and A. Malik, "Short-time energy, magnitude, zero crossing rate and autocorrelation measurement for discriminating voiced and unvoiced segments of speech signals," in *Proc. Int. Conf. Technological Adv. Elect. Electron. Comput. Eng.*, 2013, pp. 208–212.
[55] X. Guo, J. Liu, and Y. Chen, "FitCoach: Virtual fitness coach empowered by wearable mobile devices," in *Proc. IEEE Conf. Comput. Commun.*, 2017, pp. 1–9.
[56] J. Vartiainen, J. J. Lehtomaki, and H. Saarnisaari, "Double-threshold based narrowband signal extraction," in *Proc. IEEE 61st Veh. Technol. Conf.*, 2005, pp. 1–5.
[57] S. Lee et al., "An ultrathin conformable vibration-responsive electronic skin for quantitative vocal recognition," *Nature Commun.*, vol. 10, no. 1, pp. 1–11, 2019.
[58] D. M. Watt, "Gnathosonics in occlusal evaluation," *J. Prosthetic Dent.*, vol. 19, no. 2, pp. 133–143, 1968.
[59] A. M. Abduljabbar, M. E. Yavuz, F. Costen, R. Himeno, and H. Yokota, "Frequency dispersion compensation through variable window utilization in time-reversal techniques for electromagnetic waves," *IEEE Trans. Antennas Propag.*, vol. 64, no. 8, pp. 3636–3639, Aug. 2016.
[60] S. Chopra, R. Hadsell, and Y. LeCun, "Learning a similarity metric discriminatively, with application to face verification," in *Proc. IEEE Comput. Soc. Conf. Comput. Vis. Pattern Recognit.*, 2005, pp. 1–8.
[61] Y. Xie, F. Li, Y. Wu, S. Yang, and Y. Wang, "Real-time detection for drowsy driving via acoustic sensing on smartphones," *IEEE Trans. Mobile Comput.*, vol. 20, no. 8, pp. 2671–2685, Aug. 2021.
[62] WeChat, "Voiceprint: The New WeChat Password," 2015. [Online]. Available: https://blog.wechat.com/2015/05/21/voiceprint-the-new-wechat-password/
[63] X. Sun et al., "Earmonitor: In-ear motion-resilient acoustic sensing using commodity earphones," in *Proc. ACM Interactive Mobile Wearable Ubiquitous Technol.*, vol. 6, no. 4, pp. 182:1–182:22, 2022.

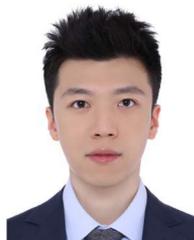

**Yadong Xie** (Graduate Student Member, IEEE) received the BE degree in network engineering from Hebei University, China, in 2016. He is currently working toward the PhD degree with the School of Computer Science, Beijing Institute of Technology, Beijing, China. His research interests include mobile computing, mobile health, human-computer interaction, and deep learning.

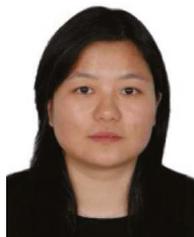

**Fan Li** (Member, IEEE) received the BEng and MEng degrees in communications and information system from the Huazhong University of Science and Technology, China, in 1998 and 2001, respectively, the MEng degree in electrical engineering from the University of Delaware, in 2004, and the PhD degree in computer science from the University of North Carolina at Charlotte, in 2008. She is currently a professor with the School of Computer Science, Beijing Institute of Technology, China. Her current research focuses on wireless networks, ad hoc and sensor networks, and mobile computing. Her papers won Best Paper Awards from IEEE MASS (2013), IEEE IPCCC (2013), ACM MobiHoc (2014), and Tsinghua Science and Technology (2015). She is a member of the ACM.

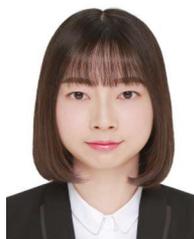

**Yue Wu** received the BE degree in Internet of things from the Beijing Institute of Technology, China, in 2015, and the PhD degree in computer science from the Beijing Institute of Technology, China, in 2021. Currently, she is a postdoctor with the School of Software, Tsinghua University, China. Her research interests include visual SLAM, event-based visual, and wireless sensor network.

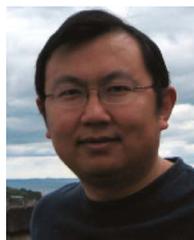

**Yu Wang** (Fellow, IEEE) received the BEng and MEng degrees in computer science from Tsinghua University, and the PhD degree in computer science from the Illinois Institute of Technology. He is currently a professor with the Department of Computer and Information Sciences, Temple University. His research interests include wireless networks, smart sensing, and mobile computing. He has published more than 200 papers in peer reviewed journals and conferences, with four best paper awards. He has served as general chair, program chair, program committee member, etc. for many international conferences (such as IEEE IPCCC, ACM MobiHoc, IEEE INFOCOM, IEEE GLOBECOM, IEEE ICC). He has served as editorial board member of several international journals, including the *IEEE Transactions on Parallel and Distributed Systems*. He is a recipient of Ralph E. Powe Junior Faculty Enhancement Awards from Oak Ridge Associated Universities (2006), Outstanding Faculty Research Award from College of Computing and Informatics, University of North Carolina at Charlotte (2008), and ACM distinguished member (2020).